\documentclass[10pt, conference, letterpaper]{IEEEtran}
\usepackage{url}
\usepackage{blindtext, graphicx}
\usepackage{relsize}
\usepackage{cite}
\usepackage{tabularx, colortbl}
\usepackage[table,xcdraw]{xcolor}
\usepackage{bm}
\usepackage{changepage}
\usepackage{subfigure}
\usepackage{anyfontsize}
\usepackage[breaklinks,colorlinks=true,citecolor=blue,pdftex]{hyperref}

\begin{document}

\title{\huge Minimizing Flow Completion Times using Adaptive Routing over Inter-Datacenter Wide Area Networks}

\author{
    \IEEEauthorblockN{Mohammad Noormohammadpour and Cauligi S. Raghavendra}
    \IEEEauthorblockA{Ming Hsieh Department of Electrical Engineering, University of Southern California (USC)}
}

\maketitle

\begin{abstract}
Inter-datacenter networks connect dozens of geographically dispersed datacenters and carry traffic flows with highly variable sizes and different classes. Adaptive flow routing can improve efficiency and performance by assigning paths to new flows according to network status and flow properties. A popular approach widely used for traffic engineering is based on current bandwidth utilization of links. We propose an alternative that reduces bandwidth usage by up to at least $50\%$ and flow completion times by up to at least $40\%$ across various scheduling policies and flow size distributions.
\end{abstract}


\section{Introduction}
Inter-datacenter networks that are managed by one organization connect dozens of geographically dispersed datacenters offering dedicated bandwidth and high visibility into network status. Traffic over these networks can generally be classified as interactive user flows which are highly sensitive to latency and internal flows generated due to internal business operations. Internal flows are less sensitive to initial routing latency, can be significantly larger than user flows, can vary by several orders of magnitude in demand (size) \cite{social_inside}, and constitute the majority of total traffic volume \cite{facebook-express-backbone}. Using rate-limiting and bandwidth allocation over inter-datacenter networks, theoretically zero and practically minimal packet loss due to congestion can be achieved \cite{tempus}. 

We focus on single path routing while aiming at minimizing completion times and bandwidth usage of internal flows. Adaptive flow routing can be used to improve performance and efficiency by assigning paths to new flows according to network status and flow properties. Although path selection can be formulated as an online optimization problem, such problems cannot be solved optimally due to limited knowledge about future flow arrivals. Alternatively, heuristic schemes can be used by considering a cost (distance) metric and selecting the minimum cost (shortest) path. 

A cost metric which has been extensively used by prior work is bandwidth utilization \cite{ospf-is_is, texcp, tempus}. We argue that while assigning paths to new flows, instead of focusing on current bandwidth utilization, \textit{one should consider utilization temporally and into the future,} i.e., by counting total outstanding bytes to be sent per link according to paths assigned to flows and total outstanding bytes per flow. We refer to this total number of remaining bytes per link as its \textit{load} and propose using it as the cost metric. Compared to utilization, load offers more information about future usage of a link's bandwidth which can help us perform more effective load balancing. Every time a flow is assigned to a path, load variables associated with all edges of that path increase by its demand. Also, a link's load variable decreases continuously as flows on that link make progress.

In addition, we evaluate two heuristics of selecting the path with minimum value of maximum link cost and minimum value of sum of link costs which we refer to as \texttt{MINMAX()} and \texttt{MINSUM()}, respectively. Although the former is frequently used in the literature \cite{ospf-is_is, texcp, tempus}, we find that the latter offers considerably better performance for the majority of traffic patterns and scheduling policies.


\begin{figure*}[t]
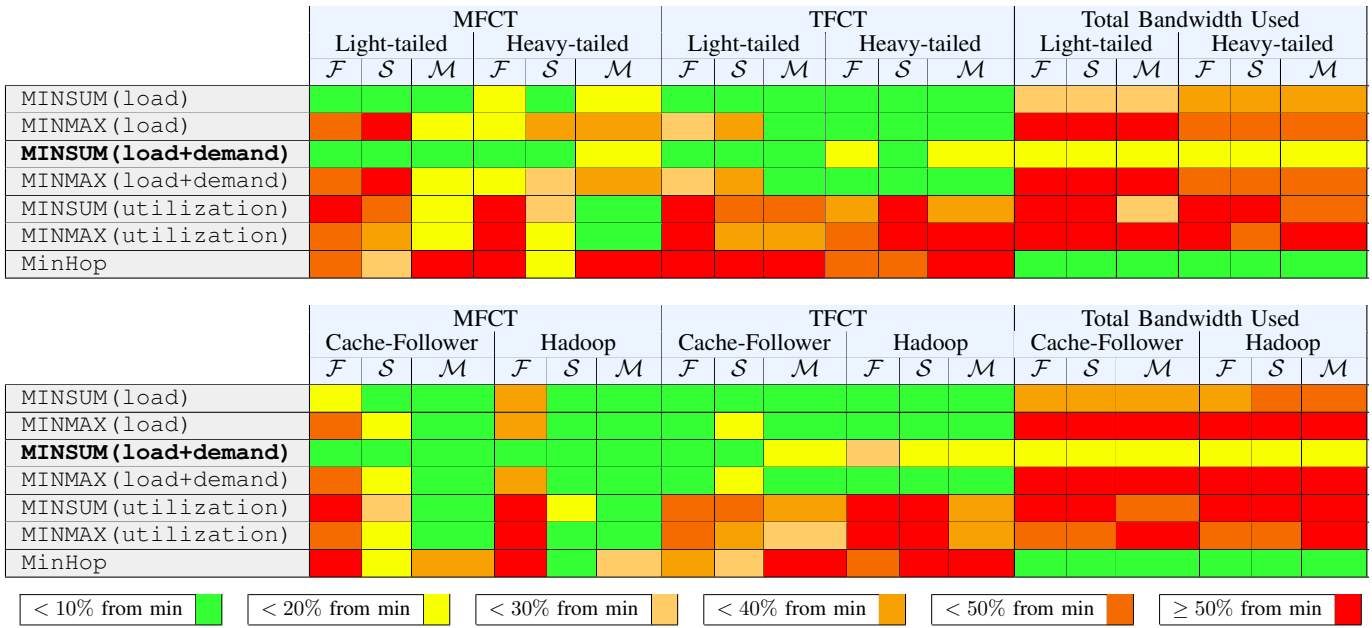

\centering
{\small
\begin{tabularx}{\textwidth}{l|l|l|l|l|l|l|l|l|l|l|l|l|l|l|l|l|l|l|}
\cline{2-19}
 & \multicolumn{6}{c|}{\cellcolor[HTML]{ECF4FF}\parbox{4.25cm}{\centering MFCT}} & \multicolumn{6}{c|}{\cellcolor[HTML]{ECF4FF}\parbox{4.25cm}{\centering TFCT}} & \multicolumn{6}{c|}{\cellcolor[HTML]{ECF4FF}\parbox{4.25cm}{\centering Total Bandwidth Used}} \\ \cline{2-19} 
 & \multicolumn{3}{c|}{\cellcolor[HTML]{ECF4FF}\small Light-tailed} & \multicolumn{3}{c|}{\cellcolor[HTML]{ECF4FF}\small Heavy-tailed} & \multicolumn{3}{c|}{\cellcolor[HTML]{ECF4FF}\small Light-tailed} & \multicolumn{3}{c|}{\cellcolor[HTML]{ECF4FF}\small Heavy-tailed} & \multicolumn{3}{c|}{\cellcolor[HTML]{ECF4FF}\small Light-tailed} & \multicolumn{3}{c|}{\cellcolor[HTML]{ECF4FF}\small Heavy-tailed} \\ \cline{2-19} 
 & \multicolumn{1}{c|}{\cellcolor[HTML]{ECF4FF}$\mathcal{F}$} & \multicolumn{1}{c|}{\cellcolor[HTML]{ECF4FF}$\mathcal{S}$} & \multicolumn{1}{c|}{\cellcolor[HTML]{ECF4FF}$\mathcal{M}$} & \multicolumn{1}{c|}{\cellcolor[HTML]{ECF4FF}$\mathcal{F}$} & \multicolumn{1}{c|}{\cellcolor[HTML]{ECF4FF}$\mathcal{S}$} & \multicolumn{1}{c|}{\cellcolor[HTML]{ECF4FF}$\mathcal{M}$} & \multicolumn{1}{c|}{\cellcolor[HTML]{ECF4FF}$\mathcal{F}$} & \multicolumn{1}{c|}{\cellcolor[HTML]{ECF4FF}$\mathcal{S}$} & \multicolumn{1}{c|}{\cellcolor[HTML]{ECF4FF}$\mathcal{M}$} & \multicolumn{1}{c|}{\cellcolor[HTML]{ECF4FF}$\mathcal{F}$} & \multicolumn{1}{c|}{\cellcolor[HTML]{ECF4FF}$\mathcal{S}$} & \multicolumn{1}{c|}{\cellcolor[HTML]{ECF4FF}$\mathcal{M}$} & \multicolumn{1}{c|}{\cellcolor[HTML]{ECF4FF}$\mathcal{F}$} & \multicolumn{1}{c|}{\cellcolor[HTML]{ECF4FF}$\mathcal{S}$} & \multicolumn{1}{c|}{\cellcolor[HTML]{ECF4FF}$\mathcal{M}$} & \multicolumn{1}{c|}{\cellcolor[HTML]{ECF4FF}$\mathcal{F}$} & \multicolumn{1}{c|}{\cellcolor[HTML]{ECF4FF}$\mathcal{S}$} & \multicolumn{1}{c|}{\cellcolor[HTML]{ECF4FF}$\mathcal{M}$} \\ \hline
\multicolumn{1}{|l|}{\cellcolor[HTML]{EFEFEF}{\color[HTML]{000000} \texttt{MINSUM(load)}}} & \cellcolor[HTML]{34FF34} & \cellcolor[HTML]{34FF34} & \cellcolor[HTML]{34FF34} & \cellcolor[HTML]{F8FF00} & \cellcolor[HTML]{34FF34} & \cellcolor[HTML]{F8FF00} & \cellcolor[HTML]{34FF34} & \cellcolor[HTML]{34FF34} & \cellcolor[HTML]{34FF34} & \cellcolor[HTML]{34FF34} & \cellcolor[HTML]{34FF34} & \cellcolor[HTML]{34FF34} & \cellcolor[HTML]{FFCC67} & \cellcolor[HTML]{FFCC67} & \cellcolor[HTML]{FFCC67} & \cellcolor[HTML]{F8A102} & \cellcolor[HTML]{F8A102} & \cellcolor[HTML]{F8A102}  \\ \hline
\multicolumn{1}{|l|}{\cellcolor[HTML]{EFEFEF}{\color[HTML]{000000} \texttt{MINMAX(load)}}} & \cellcolor[HTML]{F56B00} & \cellcolor[HTML]{FE0000} & \cellcolor[HTML]{F8FF00} & \cellcolor[HTML]{F8FF00} & \cellcolor[HTML]{F8A102} & \cellcolor[HTML]{F8A102} & \cellcolor[HTML]{FFCC67} & \cellcolor[HTML]{F8A102} & \cellcolor[HTML]{34FF34} & \cellcolor[HTML]{34FF34} & \cellcolor[HTML]{34FF34} & \cellcolor[HTML]{34FF34} & \cellcolor[HTML]{FE0000} & \cellcolor[HTML]{FE0000} & \cellcolor[HTML]{FE0000} & \cellcolor[HTML]{F56B00} & \cellcolor[HTML]{F56B00} & \cellcolor[HTML]{F56B00}  \\ \hline
\multicolumn{1}{|l|}{\cellcolor[HTML]{EFEFEF}{\color[HTML]{000000} \textbf{\texttt{MINSUM(load+demand)}}}} & \cellcolor[HTML]{34FF34} & \cellcolor[HTML]{34FF34} & \cellcolor[HTML]{34FF34} & \cellcolor[HTML]{34FF34} & \cellcolor[HTML]{34FF34} & \cellcolor[HTML]{F8FF00} & \cellcolor[HTML]{34FF34} & \cellcolor[HTML]{34FF34} & \cellcolor[HTML]{34FF34} & \cellcolor[HTML]{F8FF00} & \cellcolor[HTML]{34FF34} & \cellcolor[HTML]{F8FF00} & \cellcolor[HTML]{F8FF00} & \cellcolor[HTML]{F8FF00} & \cellcolor[HTML]{F8FF00} & \cellcolor[HTML]{F8FF00} & \cellcolor[HTML]{F8FF00} & \cellcolor[HTML]{F8FF00}  \\ \hline
\multicolumn{1}{|l|}{\cellcolor[HTML]{EFEFEF}{\color[HTML]{000000} \texttt{MINMAX(load+demand)}}} & \cellcolor[HTML]{F56B00} & \cellcolor[HTML]{FE0000} & \cellcolor[HTML]{F8FF00} & \cellcolor[HTML]{F8FF00} & \cellcolor[HTML]{FFCC67} & \cellcolor[HTML]{F8A102} & \cellcolor[HTML]{FFCC67} & \cellcolor[HTML]{F8A102} & \cellcolor[HTML]{34FF34} & \cellcolor[HTML]{34FF34} & \cellcolor[HTML]{34FF34} & \cellcolor[HTML]{34FF34} & \cellcolor[HTML]{FE0000} & \cellcolor[HTML]{FE0000} & \cellcolor[HTML]{FE0000} & \cellcolor[HTML]{F56B00} & \cellcolor[HTML]{F56B00} & \cellcolor[HTML]{F56B00}  \\ \hline
\multicolumn{1}{|l|}{\cellcolor[HTML]{EFEFEF}{\color[HTML]{000000} \texttt{MINSUM(utilization)}}} & \cellcolor[HTML]{FE0000} & \cellcolor[HTML]{F56B00} & \cellcolor[HTML]{F8FF00} & \cellcolor[HTML]{FE0000} & \cellcolor[HTML]{FFCC67} & \cellcolor[HTML]{34FF34} & \cellcolor[HTML]{FE0000} & \cellcolor[HTML]{F56B00} & \cellcolor[HTML]{F56B00} & \cellcolor[HTML]{F8A102} & \cellcolor[HTML]{FE0000} & \cellcolor[HTML]{F8A102} & \cellcolor[HTML]{FE0000} & \cellcolor[HTML]{FE0000} & \cellcolor[HTML]{FFCC67} & \cellcolor[HTML]{FE0000} & \cellcolor[HTML]{FE0000} & \cellcolor[HTML]{F56B00}  \\ \hline
\multicolumn{1}{|l|}{\cellcolor[HTML]{EFEFEF}{\color[HTML]{000000} \texttt{MINMAX(utilization)}}} & \cellcolor[HTML]{F56B00} & \cellcolor[HTML]{F8A102} & \cellcolor[HTML]{F8FF00} & \cellcolor[HTML]{FE0000} & \cellcolor[HTML]{F8FF00} & \cellcolor[HTML]{34FF34} & \cellcolor[HTML]{FE0000} & \cellcolor[HTML]{F8A102} & \cellcolor[HTML]{F8A102} & \cellcolor[HTML]{F56B00} & \cellcolor[HTML]{FE0000} & \cellcolor[HTML]{FE0000} & \cellcolor[HTML]{FE0000} & \cellcolor[HTML]{FE0000} & \cellcolor[HTML]{FE0000} & \cellcolor[HTML]{FE0000} & \cellcolor[HTML]{F56B00} & \cellcolor[HTML]{FE0000}  \\ \hline
\multicolumn{1}{|l|}{\cellcolor[HTML]{EFEFEF}{\color[HTML]{000000} \texttt{MinHop}}} & \cellcolor[HTML]{F56B00} & \cellcolor[HTML]{FFCC67} & \cellcolor[HTML]{FE0000} & \cellcolor[HTML]{FE0000} & \cellcolor[HTML]{F8FF00} & \cellcolor[HTML]{FE0000} & \cellcolor[HTML]{FE0000} & \cellcolor[HTML]{FE0000} & \cellcolor[HTML]{FE0000} & \cellcolor[HTML]{F56B00} & \cellcolor[HTML]{F56B00} & \cellcolor[HTML]{FE0000} & \cellcolor[HTML]{34FF34} & \cellcolor[HTML]{34FF34} & \cellcolor[HTML]{34FF34} & \cellcolor[HTML]{34FF34} & \cellcolor[HTML]{34FF34} & \cellcolor[HTML]{34FF34}  \\ \hline
\end{tabularx}
}

\vspace{1.0em}

{\small
\begin{tabularx}{\textwidth}{l|l|l|l|l|l|l|l|l|l|l|l|l|l|l|l|l|l|l|}
\cline{2-19}
 & \multicolumn{6}{c|}{\cellcolor[HTML]{ECF4FF}\parbox{4.25cm}{\centering MFCT}} & \multicolumn{6}{c|}{\cellcolor[HTML]{ECF4FF}\parbox{4.25cm}{\centering TFCT}} & \multicolumn{6}{c|}{\cellcolor[HTML]{ECF4FF}\parbox{4.25cm}{\centering Total Bandwidth Used}} \\ \cline{2-19} 
 & \multicolumn{3}{c|}{\cellcolor[HTML]{ECF4FF}\small Cache-Follower} & \multicolumn{3}{c|}{\cellcolor[HTML]{ECF4FF}\small Hadoop} & \multicolumn{3}{c|}{\cellcolor[HTML]{ECF4FF}\small Cache-Follower} & \multicolumn{3}{c|}{\cellcolor[HTML]{ECF4FF}\small Hadoop} & \multicolumn{3}{c|}{\cellcolor[HTML]{ECF4FF}\small Cache-Follower} & \multicolumn{3}{c|}{\cellcolor[HTML]{ECF4FF}\small Hadoop} \\ \cline{2-19} 
 & \multicolumn{1}{c|}{\cellcolor[HTML]{ECF4FF}\small $\mathcal{F}$} & \multicolumn{1}{c|}{\cellcolor[HTML]{ECF4FF}\small $\mathcal{S}$} & \multicolumn{1}{c|}{\cellcolor[HTML]{ECF4FF}\small $\mathcal{M}$} & \multicolumn{1}{c|}{\cellcolor[HTML]{ECF4FF}\small $\mathcal{F}$} & \multicolumn{1}{c|}{\cellcolor[HTML]{ECF4FF}\small $\mathcal{S}$} & \multicolumn{1}{c|}{\cellcolor[HTML]{ECF4FF}\small $\mathcal{M}$} & \multicolumn{1}{c|}{\cellcolor[HTML]{ECF4FF}\small $\mathcal{F}$} & \multicolumn{1}{c|}{\cellcolor[HTML]{ECF4FF}\small $\mathcal{S}$} & \multicolumn{1}{c|}{\cellcolor[HTML]{ECF4FF}\small $\mathcal{M}$} & \multicolumn{1}{c|}{\cellcolor[HTML]{ECF4FF}\small $\mathcal{F}$} & \multicolumn{1}{c|}{\cellcolor[HTML]{ECF4FF}\small $\mathcal{S}$} & \multicolumn{1}{c|}{\cellcolor[HTML]{ECF4FF}\small $\mathcal{M}$} & \multicolumn{1}{c|}{\cellcolor[HTML]{ECF4FF}\small $\mathcal{F}$} & \multicolumn{1}{c|}{\cellcolor[HTML]{ECF4FF}\small $\mathcal{S}$} & \multicolumn{1}{c|}{\cellcolor[HTML]{ECF4FF}\small $\mathcal{M}$} & \multicolumn{1}{c|}{\cellcolor[HTML]{ECF4FF}\small $\mathcal{F}$} & \multicolumn{1}{c|}{\cellcolor[HTML]{ECF4FF}\small $\mathcal{S}$} & \multicolumn{1}{c|}{\cellcolor[HTML]{ECF4FF}\small $\mathcal{M}$} \\ \hline
\multicolumn{1}{|l|}{\cellcolor[HTML]{EFEFEF}{\color[HTML]{000000} \texttt{MINSUM(load)}}} & \cellcolor[HTML]{F8FF00} & \cellcolor[HTML]{34FF34} & \cellcolor[HTML]{34FF34} & \cellcolor[HTML]{F8A102} & \cellcolor[HTML]{34FF34} & \cellcolor[HTML]{34FF34} & \cellcolor[HTML]{34FF34} & \cellcolor[HTML]{34FF34} & \cellcolor[HTML]{34FF34} & \cellcolor[HTML]{34FF34} & \cellcolor[HTML]{34FF34} & \cellcolor[HTML]{34FF34} & \cellcolor[HTML]{F8A102} & \cellcolor[HTML]{F8A102} & \cellcolor[HTML]{F8A102} & \cellcolor[HTML]{F8A102} & \cellcolor[HTML]{F56B00} & \cellcolor[HTML]{F56B00} \\ \hline
\multicolumn{1}{|l|}{\cellcolor[HTML]{EFEFEF}{\color[HTML]{000000} \texttt{MINMAX(load)}}} & \cellcolor[HTML]{F56B00} & \cellcolor[HTML]{F8FF00} & \cellcolor[HTML]{34FF34} & \cellcolor[HTML]{F8A102} & \cellcolor[HTML]{34FF34} & \cellcolor[HTML]{34FF34} & \cellcolor[HTML]{34FF34} & \cellcolor[HTML]{F8FF00} & \cellcolor[HTML]{34FF34} & \cellcolor[HTML]{34FF34} & \cellcolor[HTML]{34FF34} & \cellcolor[HTML]{34FF34} & \cellcolor[HTML]{FE0000} & \cellcolor[HTML]{FE0000} & \cellcolor[HTML]{FE0000} & \cellcolor[HTML]{FE0000} & \cellcolor[HTML]{FE0000} & \cellcolor[HTML]{FE0000} \\ \hline
\multicolumn{1}{|l|}{\cellcolor[HTML]{EFEFEF}{\color[HTML]{000000} \textbf{\texttt{MINSUM(load+demand)}}}} & \cellcolor[HTML]{34FF34} & \cellcolor[HTML]{34FF34} & \cellcolor[HTML]{34FF34} & \cellcolor[HTML]{34FF34} & \cellcolor[HTML]{34FF34} & \cellcolor[HTML]{34FF34} & \cellcolor[HTML]{34FF34} & \cellcolor[HTML]{34FF34} & \cellcolor[HTML]{F8FF00} & \cellcolor[HTML]{FFCC67} & \cellcolor[HTML]{F8FF00} & \cellcolor[HTML]{F8FF00} & \cellcolor[HTML]{F8FF00} & \cellcolor[HTML]{F8FF00} & \cellcolor[HTML]{F8FF00} & \cellcolor[HTML]{F8FF00} & \cellcolor[HTML]{F8FF00} & \cellcolor[HTML]{F8FF00} \\ \hline
\multicolumn{1}{|l|}{\cellcolor[HTML]{EFEFEF}{\color[HTML]{000000} \texttt{MINMAX(load+demand)}}} & \cellcolor[HTML]{F56B00} & \cellcolor[HTML]{F8FF00} & \cellcolor[HTML]{34FF34} & \cellcolor[HTML]{F8A102} & \cellcolor[HTML]{34FF34} & \cellcolor[HTML]{34FF34} & \cellcolor[HTML]{34FF34} & \cellcolor[HTML]{F8FF00} & \cellcolor[HTML]{34FF34} & \cellcolor[HTML]{34FF34} & \cellcolor[HTML]{34FF34} & \cellcolor[HTML]{34FF34} & \cellcolor[HTML]{FE0000} & \cellcolor[HTML]{FE0000} & \cellcolor[HTML]{FE0000} & \cellcolor[HTML]{FE0000} & \cellcolor[HTML]{FE0000} & \cellcolor[HTML]{FE0000} \\ \hline
\multicolumn{1}{|l|}{\cellcolor[HTML]{EFEFEF}{\color[HTML]{000000} \texttt{MINSUM(utilization)}}} & \cellcolor[HTML]{FE0000} & \cellcolor[HTML]{FFCC67} & \cellcolor[HTML]{34FF34} & \cellcolor[HTML]{FE0000} & \cellcolor[HTML]{F8FF00} & \cellcolor[HTML]{34FF34} & \cellcolor[HTML]{F56B00} & \cellcolor[HTML]{F56B00} & \cellcolor[HTML]{F8A102} & \cellcolor[HTML]{FE0000} & \cellcolor[HTML]{FE0000} & \cellcolor[HTML]{F8A102} & \cellcolor[HTML]{FE0000} & \cellcolor[HTML]{FE0000} & \cellcolor[HTML]{F56B00} & \cellcolor[HTML]{FE0000} & \cellcolor[HTML]{FE0000} & \cellcolor[HTML]{FE0000} \\ \hline
\multicolumn{1}{|l|}{\cellcolor[HTML]{EFEFEF}{\color[HTML]{000000} \texttt{MINMAX(utilization)}}} & \cellcolor[HTML]{F56B00} & \cellcolor[HTML]{F8FF00} & \cellcolor[HTML]{34FF34} & \cellcolor[HTML]{FE0000} & \cellcolor[HTML]{34FF34} & \cellcolor[HTML]{34FF34} & \cellcolor[HTML]{F56B00} & \cellcolor[HTML]{F8A102} & \cellcolor[HTML]{FFCC67} & \cellcolor[HTML]{FE0000} & \cellcolor[HTML]{FE0000} & \cellcolor[HTML]{F8A102} & \cellcolor[HTML]{F56B00} & \cellcolor[HTML]{F56B00} & \cellcolor[HTML]{FE0000} & \cellcolor[HTML]{F56B00} & \cellcolor[HTML]{F56B00} & \cellcolor[HTML]{FE0000} \\ \hline
\multicolumn{1}{|l|}{\cellcolor[HTML]{EFEFEF}{\color[HTML]{000000} \texttt{MinHop}}} & \cellcolor[HTML]{FE0000} & \cellcolor[HTML]{F8FF00} & \cellcolor[HTML]{F8A102} & \cellcolor[HTML]{FE0000} & \cellcolor[HTML]{34FF34} & \cellcolor[HTML]{FFCC67} & \cellcolor[HTML]{F8A102} & \cellcolor[HTML]{FFCC67} & \cellcolor[HTML]{FE0000} & \cellcolor[HTML]{F56B00} & \cellcolor[HTML]{FE0000} & \cellcolor[HTML]{FE0000} & \cellcolor[HTML]{34FF34} & \cellcolor[HTML]{34FF34} & \cellcolor[HTML]{34FF34} & \cellcolor[HTML]{34FF34} & \cellcolor[HTML]{34FF34} & \cellcolor[HTML]{34FF34} \\ \hline
\end{tabularx}
}

\vspace{0.2em}

\resizebox{\textwidth}{!}{
\renewcommand{\arraystretch}{1.2}
\begin{tabular}{|l|
>{\columncolor[HTML]{34FF34}}l |l|l|
>{\columncolor[HTML]{F8FF00}}l |l|l|
>{\columncolor[HTML]{FFCC67}}l |l|l|
>{\columncolor[HTML]{F8A102}}l |l|l|
>{\columncolor[HTML]{F56B00}}l |l|l|
>{\columncolor[HTML]{FE0000}}l |}
\cline{1-2} \cline{4-5} \cline{7-8} \cline{10-11} \cline{13-14} \cline{16-17}
$< 10\%$ from min &  &  & $< 20\%$ from min &  &  & $< 30\%$ from min &  &  & $< 40\%$ from min &  &  & $< 50\%$ from min &  &  & $\ge 50\%$ from min &  \\ \cline{1-2} \cline{4-5} \cline{7-8} \cline{10-11} \cline{13-14} \cline{16-17} 
\end{tabular}
}%

\vspace{0.5em}

\caption{Performance of various cost metrics for path selection over Cogent WAN \cite{cogent}, with uniform capacity of $1$ and $\lambda = 1.0$ ($\mathcal{F}$, $\mathcal{S}$ and $\mathcal{M}$ represent the FCFS, SRPT and MMF scheduling policies, respectively), simulation was repeated many times and average was computed} \label{fig:cogent}
\end{figure*}

\section{Evaluation of Different Cost Metrics}
We considered a large WAN called Cogent \cite{cogent} with $197$ nodes and $243$ links, four flow demand distributions of light-tailed (Exponential distribution), heavy-tailed (Pareto distribution), Cache-Follower \cite{social_inside} and Hadoop \cite{social_inside} (the last two happen across Facebook datacenters), and a uniform capacity of $1.0$ for all links. A Poisson distribution with rate $\lambda$ was used for flow arrivals. For all flow demand distributions, we assumed an average of $20$ units and a maximum of $500$ units. For heavy-tailed, we used a minimum demand of $2$ units. We considered scheduling policies of First Come First Serve (FCFS), Shortest Remaining Processing Time (SRPT) and Fair Sharing using Max-Min Fairness (MMF). We considered three different cost metrics of ``utilization", ``load", and ``load+demand" per link where demand represents the new flow's size in bytes. To measure a path's cost, we considered two cost functions of \textit{maximum} which assigns any path the cost of its highest cost link (used by \texttt{MINMAX()} heuristic), and \textit{sum} which computes a path's cost by summing up costs of its links (used by \texttt{MINSUM()} heuristic). Combining these path cost functions with the three link cost metrics mentioned above, we obtain six different path selection schemes that select the path with minimum cost for a newly arriving flow. We also considered \texttt{MinHop} which selects a path with minimum hops per flow to compute lower bound of bandwidth usage. For minimum cost path selection, we used Dijkstra's algorithm in JGraphT library. We measured Mean and Tail Flow Completion Times (MFCT/TFCT) and total bandwidth as shown in Figure \ref{fig:cogent}.

\textit{Flow Completion Times (FCT):} \texttt{MINSUM(load)} and \texttt{MINSUM(load+demand)} perform almost identically in completion times. The rest of schemes offer highly varying performance dictated by scheduling policy or traffic pattern. Schemes based on utilization are at least $40\%$ above the minimum for the majority of scenarios. Also, \texttt{MINMAX(load)} and \texttt{MINMAX(load+demand)} are more than $50\%$ above the minimum in mean completion times for multiple scenarios. Overall, it can be seen that schemes based on ``load" as link cost offer much better tail completion times (less than $10\%$ away from minimum for majority of cases). Also, \texttt{MINSUM(load+demand)} offers the best mean completion times considering all scenarios.

\textit{Total Bandwidth Usage:} \texttt{MINSUM(load+demand)} offers the minimum extra bandwidth usage compared to \texttt{MinHop} which is below $20\%$ at all times. Schemes based on \texttt{MINMAX()} consume at least $40\%$ extra bandwidth. \texttt{MINSUM(load)} and \texttt{MINSUM(utilization)} use at least $10\%$ more bandwidth at all times compared to \texttt{MINSUM(load+demand)} and at least $20\%$ more bandwidth for the majority of scenarios.

\section{Conclusions and Future Work}
We see that \texttt{MINSUM(load+demand)} stays within $20\%$ of minimum for all completion times and within $10\%$ of minimum in the majority of cases. It offers the minimum bandwidth usage across all adaptive approaches (\texttt{MinHop} is static). With this cost metric, larger flows are most likely assigned shorter paths which allows for higher bandwidth savings (due to presence of ``demand" as part of link cost) while shorter flows are assigned to paths with smaller total load which reduces completion times via load balancing. We believe \texttt{MINSUM(load+demand)} performs better than techniques based on \texttt{MINMAX()} since it considers total number of bytes that will eventually be scheduled on a path taking into account all edges and not just the highest loaded/utilized link. Our experiments have shown that \texttt{MINSUM(load+demand)} is also an effective metric for selection of multicast forwarding trees that reduce completion times via load balancing \cite{dccast, quickcast}. It is also interesting to note that \texttt{MINMAX(utilization)}, which is frequently used in traffic engineering research, is far from the best solution for the majority of evaluated scenarios. 

Centralized frameworks, such as SDN \cite{sdn}, are good candidates for realization of this scheme since they offer access to global view of network status and flow demands. To properly update load variables associated with links, one needs knowledge of flow demands. In case exact flow size is unknown, an estimate can be used. Further research is needed on how flow demand estimation accuracy can affect quality of selected paths. In addition, we plan to extend and evaluate our proposed adaptive approach for multipath traffic engineering.

{\bibliography{citations.bib}

\begin{thebibliography}{1}
\providecommand{\url}[1]{#1}
\csname url@samestyle\endcsname
\providecommand{\newblock}{\relax}
\providecommand{\bibinfo}[2]{#2}
\providecommand{\BIBentrySTDinterwordspacing}{\spaceskip=0pt\relax}
\providecommand{\BIBentryALTinterwordstretchfactor}{4}
\providecommand{\BIBentryALTinterwordspacing}{\spaceskip=\fontdimen2\font plus
\BIBentryALTinterwordstretchfactor\fontdimen3\font minus
  \fontdimen4\font\relax}
\providecommand{\BIBforeignlanguage}[2]{{%
\expandafter\ifx\csname l@#1\endcsname\relax
\typeout{** WARNING: IEEEtran.bst: No hyphenation pattern has been}%
\typeout{** loaded for the language `#1'. Using the pattern for}%
\typeout{** the default language instead.}%
\else
\language=\csname l@#1\endcsname
\fi
#2}}
\providecommand{\BIBdecl}{\relax}
\BIBdecl

\bibitem{social_inside}
A.~Roy, H.~Zeng, J.~Bagga, G.~Porter, and A.~C. Snoeren, ``{Inside the Social
  Network's (Datacenter) Network},'' \emph{SIGCOMM}, pp. 123--137, 2015.

\bibitem{facebook-express-backbone}
\BIBentryALTinterwordspacing
Building express backbone: Facebook’s new long-haul network. [Online].
  Available:
  \url{https://code.facebook.com/posts/1782709872057497/building-express-backbone-facebook-s-new-long-haul-network/}
\BIBentrySTDinterwordspacing

\bibitem{tempus}
S.~Kandula, I.~Menache, R.~Schwartz, and S.~R. Babbula, ``Calendaring for wide
  area networks,'' \emph{SIGCOMM}, vol.~44, no.~4, pp. 515--526, 2015.

\bibitem{ospf-is_is}
B.~Fortz and M.~Thorup, ``{Optimizing OSPF/IS-IS weights in a changing
  world},'' \emph{IEEE Journal on Selected Areas in Communications}, vol.~20,
  no.~4, pp. 756--767, 2002.

\bibitem{texcp}
S.~Kandula \emph{et~al.}, ``{Walking the Tightrope: Responsive Yet Stable
  Traffic Engineering},'' \emph{SIGCOMM}, vol.~35, no.~4, pp. 253--264, 2005.

\bibitem{cogent}
\BIBentryALTinterwordspacing
The internet topology zoo (cogent). [Online]. Available:
  \url{http://www.topology-zoo.org/files/Cogentco.gml}
\BIBentrySTDinterwordspacing

\bibitem{dccast}
M.~Noormohammadpour, C.~S. Raghavendra, S.~Rao, and S.~Kandula, ``Dccast:
  Efficient point to multipoint transfers across datacenters,'' in
  \emph{HotCloud}.\hskip 1em plus 0.5em minus 0.4em\relax USENIX Association,
  2017.

\bibitem{quickcast}
M.~Noormohammadpour, C.~S. Raghavendra, S.~Kandula, and S.~Rao, ``{QuickCast:
  Fast and Efficient Inter-Datacenter Transfers using Forwarding Tree
  Cohorts},'' \emph{arXiv preprint arXiv:1801.00837}, 2018.

\bibitem{sdn}
N.~McKeown, ``Software-defined networking,'' \emph{INFOCOM keynote talk},
  vol.~17, no.~2, pp. 30--32, 2009.

\end{thebibliography}
\bibliographystyle{IEEEtran}}
\end{document}